\documentclass[%
reprint,
superscriptaddress,
longbibliography,
nofootinbib,
amsmath,amssymb,
floatfix,
aip,
author-year,
]{revtex4-2}

\usepackage[utf8]{inputenc}
\usepackage{amsthm,amsmath,amssymb}
\usepackage{graphicx}
\usepackage[table]{xcolor}
\definecolor{Set1-blue}{RGB}{55,126,184}
\usepackage[colorlinks=true,%
 linkcolor = {Set1-blue},%
 citecolor = {Set1-blue},%
 urlcolor = {Set1-blue}]{hyperref}

\graphicspath{ {figures/} }

\DeclareMathOperator{\lognormal}{lognormal}
\DeclareMathOperator{\logistic}{logistic}

\usepackage{booktabs}
\usepackage{dcolumn}

\begin{document}

\title{Metrics and peer review agreement at the institutional level}

\author{V.A. Traag}
\email{v.a.traag@cwts.leidenuniv.nl}
\affiliation{Centre for Science and Technology Studies (CWTS), Leiden University, the Netherlands}

\author{M. Malgarini}
\affiliation{Agenzia Nazionale di Valutazione del sistema Universitario e della Ricerca, Rome, Italy}
\author{S. Sarlo}
\affiliation{Agenzia Nazionale di Valutazione del sistema Universitario e della Ricerca, Rome, Italy}

\date{\today}

\begin{abstract}
In the past decades, many countries have started to fund academic institutions based on the evaluation of their scientific performance.
In this context, post-publication peer review is often used to assess scientific performance.
Bibliometric indicators have been suggested as an alternative to peer review.
A recurrent question in this context is whether peer review and metrics tend to yield similar outcomes.
In this paper, we study the agreement between bibliometric indicators and peer review based on a sample of publications submitted for evaluation to the national Italian research assessment exercise (2011--2014).
In particular, we study the agreement between bibliometric indicators and peer review at a higher aggregation level, namely the institutional level.
Additionally, we also quantify the internal agreement of peer review at the institutional level.
We base our analysis on a hierarchical Bayesian model using cross-validation.
We find that the level of agreement is generally higher at the institutional level than at the publication level.
Overall, the agreement between metrics and peer review is on par with the internal agreement among two reviewers for certain fields of science in this particular context.
This suggests that for some fields, bibliometric indicators may possibly be considered as an alternative to peer review for the Italian national research assessment exercise.
Although results do not necessarily generalise to other contexts, it does raise the question whether similar findings would obtain for other research assessment exercises, such as in the United Kingdom.
\end{abstract}

\keywords{research assessment; research evaluation; bibliometrics; peer review}

\maketitle

\section{Introduction}

\noindent Since the 1980s, performance-based research funding systems (PBRFS) were introduced in many countries in order to strengthen accountability of research institutions and steer their behaviour.
PBRFS may vary considerably in how they function \citep{Hicks2012,Zacharewicz2018}, but they have one element in common: the need to evaluate research.
Peer review is often considered the principal method for evaluating scientific products.
Indeed, some countries, such as the UK, have opted for research assessment that is primarily based on peer review.
In large research assessment exercises, peer review may become costly.
To facilitate the assessment, bibliometric indicators can be used to inform the judgement of peers.
In Italy, the research assessment exercise, known as the \textit{Valutazione della Qualità della Ricerca} (VQR), uses an informed peer review approach, where review by selected panellists and external peers is supported by bibliometrics in fields for which bibliometric indicators seem informative (see \citet{Ancaiani2015} for more details).

A recurrent question in this context is whether peer review and metrics tend to yield similar outcomes, or whether they differ substantially \citep{Narin1976}.
This question has been repeatedly addressed in the context of the UK Research Excellence Framework (REF), culminating in a systematic large-scale comparison between peer review and metrics in the Metric Tide report \cite[Supplementary Report II]{Wilsdon2015}.
We believe that this report has two crucial shortcomings \citep{Traag2019}: (1) the agreement between peer review and metrics was studied at the publication level, in contrast to the aggregate institutional level at which the REF outcomes are relevant; and (2) the internal agreement of peer review itself (i.e.
the extent to which different reviewers or different peer review panels come to the same conclusion) was not considered.
In the Italian context, the \emph{Agenzia Nazionale di Valutazione
del sistema Universitario e della Ricerca} (ANVUR), the agency tasked with the implementation of the VQR, collected data on peer review and quantified the internal agreement of peer review at the publication level.
The results of the analysis showed that the agreement between metrics and peer review is similar to, or higher than, the agreement between two independent reviewers \citep{Ancaiani2015,Bertocchi2015,Alfo2017}, although this result has been subject to debate \citep{Baccini2016,Baccini2017}.
In this paper, we use the ANVUR dataset to study the internal peer review agreement at the institutional level.
Similar to \citet{Traag2019} we find that the agreement between peer review and metrics tends to be higher at the institutional level than at the individual publication level.
In addition to \citet{Traag2019}, we quantify the internal peer review agreement at the institutional level, which is also higher than at the publication level.
Most importantly, we find that the agreement between metric and peer review is generally on par with the internal agreement among two reviewers for the fields included in our analysis.

In the next section, we provide a short background of PBRFS and a brief description of the Italian VQR exercise.
We then present the collected data and outline our methodology, followed by a summary of the main results obtained in our analysis.
Finally, in the conclusion, we discuss the connection with broader questions around evaluation.

\section{Performance-Based Research Funding Systems}

\noindent Since the early 1980s, public management changed around the world.
Reforms led to the redesign of the main public administration mechanisms at all levels and in all sectors, including higher education systems and public research organizations.
Over the last decades, a considerable number of countries, particularly EU member states, implemented performance-based research funding systems (PBRFS).
According to the definition provided by \citet{Hicks2012}, and used in \citet{Zacharewicz2018}, the main characteristics of PBRFS are the following:

\begin{itemize}
 \item Research output and/or impact is evaluated ex-post.
 \item The allocation of research funding depends (partly) on the outcome of the evaluation.
 \item The assessment and funding allocation takes place at the organisational level.
 \item PBRFS are a national or regional system.
\end{itemize}

PBRFS exclude any kind of degree programmes and teaching assessment.
Grant based funding, which is based on ex-ante evaluation of grant proposals, is excluded, and so are funding systems that assign funds only on the basis of the number of researchers or PhD students.
Furthermore, PBRFS provide, directly or indirectly, tools and mechanisms to allocate research funds.
They are performed at the national level, not on the local or institutional level, and they typically do not result in mere suggestions or recommendations to evaluated organisations, but affect the allocation of resources \citep{Hicks2012,Zacharewicz2018}.

Many countries have no research performance-based elements in their funding allocation at all (for a detailed examination see the work of \citet{Zacharewicz2018}).
Countries that do base their funding allocation partly on research performance do so in a variety of ways.
This includes countries such as the United Kingdom and Italy, which both implemented a PBRFS, but using a different approach.
The rationales of both exercises may be summarised as follows: (1) steering public funds allocation on the basis of quality, excellence or meritocratic criteria; (2) providing comparative information on institutions for benchmarking purposes; and (3) providing accountability regarding the effectiveness of research management and its impact in terms of public benefits \citep{Abramo2015,Franceschini2017}.
The United Kingdom developed the oldest and best-known assessment exercise, which is nowadays called the Research Excellence Framework (REF), which is a point of reference for many PBRFS.
Italy introduced a PBRFS more recently, nowadays called the \textit{Valutazione della Qualità della Ricerca} (VQR), which was partly inspired by the UK REF, although there are also marked differences.

Whereas the UK REF is believed to use metrics moderately and is mainly based on peer review assessments, the Italian VQR makes more extensive use of bibliometric indicators, especially in the STEM and Life Science sectors \citep{Zacharewicz2018}.
These differences are not absolute, but gradual: both rely partly on metrics and partly on peer review, but they do so in different ways.
Peer review is often seen as a kind of gold standard for research evaluation \citep{Wilsdon2015}.
The research community has long debated whether to use metrics or peer review for research evaluation: both methods have strengths and weaknesses.
It is in this context that we study the agreement between peer review and metrics.

\subsection{The VQR exercises}

Italy introduced a national research assessment exercise in 2006 as the \textit{Valutazione Triennale della Ricerca} (VTR), which looked back at the period 2001--2003 \cite[see][]{EUCommission2018}.
The second assessment exercise, the so-called \textit{Valutazione della Qualità della Ricerca} (VQR), the evaluation of research quality, looked back at the years 2004--2010 and its results were published in July 2013 by a new national agency, \textit{Agenzia Nazionale per la Valutazione dell'Università e della Ricerca} (ANVUR).
The third research assessment exercise (VQR 2011--2014) started in 2015 with reference to the period 2011--2014 and its results were published in February 2017 by ANVUR.
VTR and VQR results have been used by the government to allocate a growing share of the Ordinary University Fund (FFO), starting from 2.2\% of total funding in 2009 and reaching 28\% in 2019.

VQR evaluates research outputs of all permanent scientific staff in 96 universities and 39 public research organisations.
With reference to the period 2011--2014, these organisations submitted what they considered to be the best ouputs for evaluation on behalf of $52\,677$ researchers (2 for each university researcher, and 3 for each scientist employed in a public research organisation).
All in all, $118\,036$ outputs were submitted, out of which 78$\%$ were journal articles.
The remaining types of research outputs included a wide range of materials such as book chapters, conference proceedings and even works of art.
Outputs were classified in 16 research areas and ANVUR appointed a \emph{Gruppo di Esperti della Valutazione} (GEV), a panel of experts, for each research area.

In humanities and social sciences (with the exception of Psychology and Economics \& Statistics), a pure peer review system was employed, assisted by external (national and international) reviewers.
In total, for all research areas, almost $17\,000$ reviewers were used in VQR 2011--2014.
External reviewers rated the publication with a score 1--10 on three criteria: originality, methodological rigour and impact.

Generally speaking, in Science, Technology, Engineering and Medicine areas (STEM), the same review procedure was used, but in addition bibliometric indicators were also produced by ANVUR to inform the panels.
In Mathematics and Economics \& Statistics bibliometric indicators played a central role even if these two GEVs adopted slightly different bibliometric evaluation algorithms with respect to STEM and Psychology (for details see the GEV1 and GEV13 Area Reports).
In particular, in Economics \& Statistics, no bibliometric scores were gathered by ANVUR, and therefore, these variables are missing for all publications in Economics \& Statistics.
More specifically, peer evaluation was integrated with the use of bibliometric indicators concerning citations and journal impact, drawn from the major international databases (see \citet{Anfossi2016}, for a description of the bibliometric algorithm used in the exercise).
The indicators used were the 5-Year Impact Factor and the Article Influence Score (AIS) for the WoS database and the Scimago Journal Rank (SJR) and the Impact Per Publication (IPP) indicators in Scopus.
On the basis of the bibliometric algorithm, when citations and journal impact indicators provided contrasting results the paper was not assigned an evaluation class but was rather sent to external review for informed review (IR).
More specifically, if a research product was published in a high-impact journal but received few citations, or vice-versa was published in a low-impact journal but was cited frequently, the product was evaluated with peer review\footnote{See the official document of GEV2 - Physics for more details, similar procedures were adopted in the other scientific areas \url{https://www.anvur.it/wp-content/uploads/2016/02/Criteria\%20GEV\%2002_English.pdf}}.

\section{Data and indicators}

\begin{figure}
 \begin{center}
 \includegraphics[width=\linewidth]{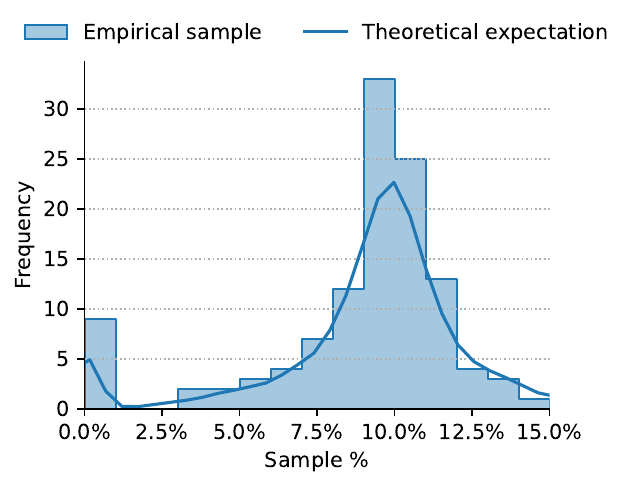}
 \end{center}
 \caption{Distribution of percentage of publications in the sample across institutions.
  The $x$-axis shows the percentage of publications of all submitted publications that are included in the sample used in our study.
  The $y$-axis shows the number of institutions with the indicated percentage of publications.
  The line shows the expected distribution of percentages based on 1000 random stratified samples.
 }
 \label{fig:stratification}
\end{figure}

\noindent The analysis carried out in this paper is based on reviews collected previously by ANVUR of a sample extracted from the full dataset of journal articles that were submitted for evaluation for VQR 2011--2014 \cite[p. 17 and Appendix B]{ANVUR2017}.
The sample was limited to the research areas that made use of metrics, i.e. all STEM areas, Psychology, and Economics \& Statistics\footnote{The excluded research areas were:
(8a) Architecture;
(10) Ancient History, Philology, Literature \& Art History;
(11a) History, Philosophy, Pedagogy;
(12) Law; and
(14) Political \& Social Sciences.
}
In total, $77\,159$ journals articles were submitted in these research areas that were evaluated through bibliometrics in VQR 2011--2014.
A random sample of 10\% of these $77\,159$ journal articles was drawn, stratified by research area, resulting in $7\,667$ sampled journal articles\footnote{Note that due to some earlier misclassifications of submissions, the sample is slightly less than 10\%.}.
See Table~\ref{tab:GEV} for an overview of the research areas and the number of articles per research area.
ANVUR sent out all journal articles in the sample to two independent peer reviewers, irrespective of their bibliometric indicators.
ANVUR did not provide reviewers with bibliometric indicators of the articles that were requested to be reviewed.
The response rate of reviewers has been high and the number of articles that were peer reviewed by two reviewers covered $7\,164$ articles.
Overall, the empirical sample could be considered a reasonable approximation of the population of reference \cite[see also][]{Alfo2017}.
Post-stratification analysis suggests that the sample is also sufficiently representative at the institutional level (Fig.~\ref{fig:stratification}).

\begin{table*}
  \begin{tabular*}{10.5cm}{llrr}
    \toprule
    ID                        & Name                                  & Submitted articles & In sample \\
    \midrule
    1                         & Mathematics \& Computer Sciences      &  4\,631               &    468 \\
    2                         & Physics                               & 10\,182               & 1\,018 \\
    3                         & Chemistry                             &  6\,625               &    662 \\
    4                         & Earth Sciences                        &  3\,953               &    394 \\
    5                         & Biology                               & 10\,423               & 1\,037 \\
    6                         & Medicine                              & 15\,400               & 1\,524 \\
    7                         & Agricultural \& veterinary sciences   &  6\,354               &    638 \\
    8b                        & Civil Engineering                     &  2\,370               &    237 \\
    9                         & Industrial \& Information Engineering &  9\,930               &    998 \\
    11b                       & Psychology                            &  1\,801               &    180 \\
    13                        & Economics \& Statistics               &  5\,490               &    511 \\
    \midrule
                                            \multicolumn{2}{r}{Total} & 77\,159               & 7\,667 \\
    \bottomrule
  \end{tabular*}
  \caption{Overview of research areas, \emph{Gruppo di Esperti della Valutazione} (GEV), and the number of submitted articles and the number of articles included in the sample that is analysed.}
  \label{tab:GEV}
\end{table*}

We matched all publications in the sample with the CWTS in-house version of Web of Science (WoS).
In total, $6\,337$ publications could be matched to WoS, $6\,001$ of which were reviewed by two reviewers, comprising $7.8\%$ of the reference population, from 110 Italian institutions.
A few institutions that only had one or two publications in the sample had no match with WoS at all.
It is worth noting that institutions were not always included in all areas.

As stated, publications in the sample were assessed by two independent reviewers.
We randomly determined which reviewer is considered as reviewer number $1$ and which one is considered as reviewer number $2$.
We summed the three scores on the three different criteria of originality, methodological rigour and impact to obtain an overall score that ranged from 3 to 30.

\begin{figure}
  \centering
  \includegraphics[width=\linewidth]{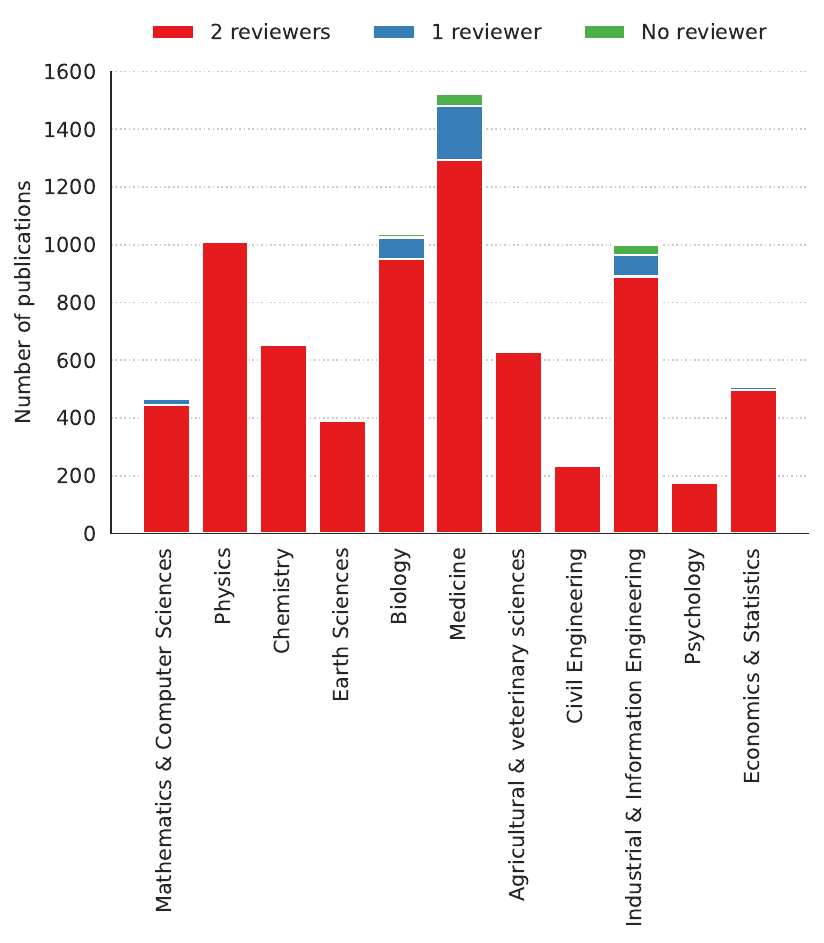}
  \caption{Number of publications in the sample with missing reviewer scores per GEV.}
  \label{fig:missing_pubs}
\end{figure}

As said, $7\,164$ publications were reviewed by two independent reviewers.
There are $122$ publications without any reviewer, and $381$ publications that were reviewed by only a single reviewer.
Many of these publications with missing reviewer scores are concentrated in Medicine, Biology and Industrial \& Information Engineering (Fig.~\ref{fig:missing_pubs}).
In our analysis, we include all publications, also those for which we miss reviewer scores or citation scores, which we explain in more detail in the methodology section.

For each paper included in the sample, we calculated two indicators on the basis of WoS: one at the article level and one at the journal level.
We calculated (1) the normalised citation score (NCS) for each paper, given by the number of citations divided by the average number of citations of all publications in the same field and the same year; and (2) the normalised journal score (NJS), which is the average NCS of all publications in a certain journal and a certain year.
We took into account citations up to (and including) 2015, to be consistent with the timing of the VQR.
We used the WoS journal subject categories for calculating normalised indicators.
In the case of journals that were assigned to multiple subject categories, we applied a fractionalisation approach to normalise citations \citep{Waltman2011a}.
Publications within journals in the multidisciplinary category (e.g. Science, Nature, PLOS ONE) were fractionally reassigned to other subject categories based on their references.
Besides the bibliometric information calculated from WoS, we also considered the indicators gathered by ANVUR during the VQR itself.
Two different types of indicators were collected: one citation-based indicator and one journal-based indicator.
Those indicators may come from various sources (e.g. Scopus, WoS, MathSciNet), and for different publications different journal indicators may be used, such as the 5-year Impact Factor, Article Influence Score, SJR, and IPP\footnote{In Mathematics \& Computer Science, the MCQ indicator extracted from the MathSciNet database was also used for a limited number of papers.}
Institutions could choose the source that should be used for the indicators.
All VQR scores were normalised as percentiles with respect to the field definitions as provided by the data source.
This procedure allowed the VQR to gain a greater degree of flexibility in practice \citep{Anfossi2016}, but also made the data more heterogeneous, thereby complicating the interpretation of the results.
Nonetheless, we also included the VQR indicators in our analysis in order to compare them to the bibliometric information obtained exclusively from WoS.
Besides the two reviewers' scores for each paper, we hence obtained the two VQR percentile metrics (a citation metric and a journal metric), and the two WoS metrics (a citation metric and a journal metric).

We want to compare the agreement between metrics and peer review in a fair way to the internal agreement of peer review.
In order to do so, we consistently compare all scores and metrics to the overall score of reviewer 2.
Internal peer review agreement is then quantified by the agreement of the score of reviewer 2 with the predicted score based on the score of reviewer 1.
Likewise, for each metric, we calculate the agreement of reviewer 2 with the predicted score based on metrics.
By performing the analysis in this way, the agreement between metrics and peer review can be compared in a fair way to the internal agreement of peer review.
If we had chosen to compare each metric to the average score of reviewers 1 and 2, this would have already cancelled out some differences in the scores of the reviewers, and as a result, the agreement of the predicted scores based on metrics with the reviewer scores would not have been directly comparable to the internal peer review agreement.

At the level of institutions, there are two views of the aggregate scores: a size-dependent view, considering the total over a certain score, and a size-independent view, considering the average over a certain score.
For the size-dependent view, we simply take the sum of each score, while for the size-independent view we take the average of each score.
Note that at the aggregate level we still speak of the score of reviewer 1 and reviewer 2, even though this refers to two sets of reviewers, not to two individual reviewers.

\section{Methodology}

\begin{figure}
  \includegraphics[width=\linewidth]{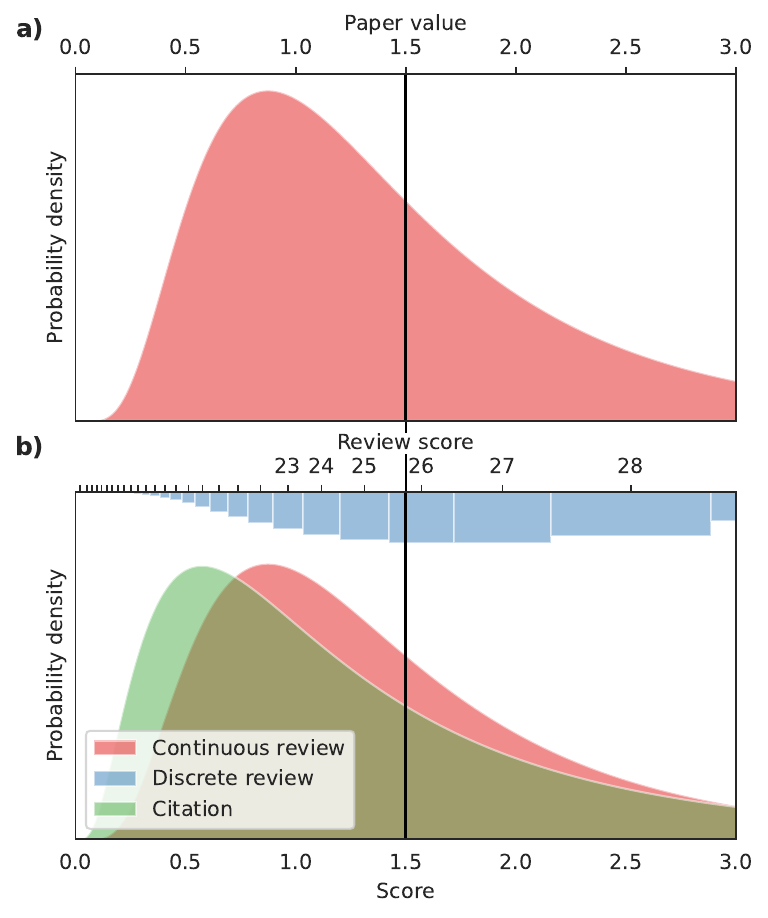}
  \caption{Illustration of the hierarchical Bayesian model.
  In (a) we show the distribution of paper values for a fictitious institution, which is distributed as $\lognormal(0.8, 0.4)$, with the solid black line representing one particular paper value of $\phi_p = 1.5$.
  In (b) we illustrate the distributions of the (non-zero) citation and review scores for the paper value $\phi_p = 1.5$ with $\sigma_\text{review} = 0.6$ and $\sigma_\text{citation} = 0.8$.
  The top axis in (b) shows the distribution of the discrete review scores $3$--$30$ corresponding with the continuous review scores shown at the bottom.}
  \label{fig:model}
\end{figure}

\noindent Following \citet{Traag2019} we do not calculate correlations, but absolute differences from predicted values.
Whereas \citet{Traag2019} focused on predicting whether a publication would be classified as $4^*$ (``World Leading''), based on whether a publication belonged to the top $10\%$ of its field, we here focus on predicting the numerical reviewer score based on the number of (normalised) citations or journal impact.
We use a hierarchical Bayesian model to predict review scores, which is similar in spirit to the simulation analysis performed by \citet{Traag2019}.
See Fig.~\ref{fig:model} for an illustration of the model.
We assume each institution $i$ has a certain institutional ``value'', $\lambda_i$, with individual papers $p$ of institution $i_p$ having a ``value'' $\phi_p$ that is distributed as
\begin{equation}
  \phi_p \sim \lognormal(\lambda_{i_p}, \sigma_\text{value}),
\end{equation}
where $\sigma_\text{value}$ reflects how broadly distributed paper values are.
Both review scores and citation scores are then assumed to be a result of the paper value $\phi_p$, where we expect higher paper values to correspond to higher review scores and citation scores.

We assume that a citation score $c_p$ of paper $p$ is generated through a lognormal hurdle model.
In our model, we assume that a citation score reflects a scaled paper value multiplied with some ``error'', that is, $c_p = \exp(\beta) \phi_p \epsilon_\text{citation}$, where $\epsilon_\text{citation}$ is the ``error''.
We assume that this ``error'' is distributed lognormal, with a mean of $1$ and some $\sigma_\text{citation}$:
\begin{equation}
  \epsilon_\text{citation} \sim \lognormal\left(-\frac{\sigma^2_\text{citation}}{2}, \sigma_\text{citation}\right).
\end{equation}
If $\sigma_\text{citation}$ is high, citation scores $c_p$ will show some variation around the paper value $\phi_p$.
Even if two papers have the same value $\phi_p$, we may observe quite different citation scores in our model.
The parameter $\beta$ reflects the relative differences of citation scores and review scores.
For $\beta < 0$, so that $\exp(\beta) < 1$, citation scores are generally lower compared to review scores for the same paper value $\phi_p$.
For $\beta > 0$, so that $\exp(\beta) > 1$, citation scores are generally higher compared to review scores for the same paper value $\phi_p$.

One complication is that we have to handle citation scores of $0$, which is not possible in a lognormal distribution.
We therefore separately estimate a probability $\Pr(c_{\bar{0}})$ that a citation score is larger than $0$, for which we use a logistic prediction
\begin{equation}
  \Pr(c_{\bar{0}}) = \logistic(\alpha_{\bar{0}} + \beta_{\bar{0}} \phi_p).
\end{equation}
Putting this together we arrive at the lognormal hurdle model
\begin{equation}
  c_p =
  \begin{cases}
    \beta \phi_p \epsilon & \text{with prob.} \Pr(c_{\bar{0}}) \\
    0 & \text{otherwise}
  \end{cases}.
\end{equation}
Based on prior literature, we expect the distribution of normalised citation scores of all of science to follow roughly a lognormal distribution with a mean of $1$ and $\sigma \approx 1.3$~\citep{Radicchi2008-ju}.
We also use this to transform percentile indicators to normalised citation scores that can be used in this model.

For a review score $r_p$ of paper $p$, we use an ordered model.
The observed review score is an integer value that ranges from 3--30, and we assume this reflects a continuous underlying evaluation score $e_p = \phi_p \epsilon_\text{review}$ where
\begin{equation}
  \epsilon_\text{review} \sim \lognormal\left(-\frac{\sigma^2_\text{review}}{2}, \sigma_\text{review}\right),
\end{equation}
where $\sigma^2_\text{review}$ captures reviewer uncertainty.
If $\sigma_\text{review}$ is high, the continuous evaluation score $e_p$ will show more variation around $\phi_p$, while for $\sigma_\text{review} = 0$, $\epsilon_\text{review} = 1$, and hence $e_p = \phi_p$, and there is no reviewer uncertainty.
We assume that continuous review scores in all of science are distributed with a mean of $1$ and a $\sigma \approx 1.3$, similar to the citation distribution.
We define thresholds $R_4$, \ldots, $R_{30}$, such that the probability for a variable $x \sim \lognormal(-1.3^2 / 2, 1.3)$ the probability to fall between $R_k$ and $R_{k+1}$ equals $\frac{1}{30 - 3 + 1}$.
In other words, we assume that the observed integer score reflects a sort of ``belonging to the top $x$\%''.
So, if the continuoues evaluation score $e_p$ falls in the interval $[R_k, R_{k+1})$ the observed review score will be $k$, and we use the overall boundaries $R_3 = 0$ and $R_{30} = \infty$.
This defines a probability to observe review score $r_p$, based on the paper value $\phi_p$.
Note that there may possibly be multiple review scores for the same paper, so that we may have $r_{p,1}$, $r_{p,2}$, etc.
Each review score is assumed to follow the same distribution as laid out here.
For all parameters we use a standard (half) normal as a prior.
Prior predictive checks showed that these priors gave reasonable results.

Our hierarchical Bayesian model naturally incorporates contextual information from an institution.
For example, suppose we have two papers of an institution, only one of which has a review score.
Based on the review score $r_1$ of paper $1$, a paper value $\phi_1$ is inferred, which hence also leads to an inferred institutional value $\lambda$.
If we are then asked to predict the review score $r_2$ of paper $2$, we then sample a paper value from the distribution based on $\lambda$.
Hence, if we observe a higher review score $r_1$ of a paper from some institution, we are more likely to predict higher review scores $r_2$ for some other paper.
This is similar to the approach taken in \citet{Traag2019}.

Our hierarchical Bayesian model naturally handles missing values.
We use the observed review scores $r_p$ and citation scores $c_p$ to infer paper values.
As said, the paper value is assumed to be drawn from a distribution of paper values that is specific to each institution.
In the absence of some observed score, whether review or citation, any other score will still help infer paper values.
If no scores are observed at all for a particular paper, the paper value is simply distributed according to the overall distribution of paper values at the institutional level.

We use $k$-fold cross-validation to separate our data into training and test sets, using $k = 5$ folds, for each metric and each research area (i.e. GEV) separately.
All data is included as test data in exactly one fold, and is used as training data in the remaing $k - 1 = 4$ folds.
It is important that we consider the hierarchical structure of the model in how the folds are defined.
In particular, institutions should be in its entirety in the training set or the test set, but not partly.
Otherwise, the model would learn already the institutional value from some publications that are in the training set, which would help the prediction in the test set, which is sometimes called leakage.
That is, if this model would be applied in practice, it could only use the general information from the parameters $\alpha$, $\beta$ and $\sigma$, but not any information from particular institutions $\lambda_i$ or papers $\phi_p$.
For the training set we use data from citations and both reviewers simultaneously, while we only use data from either reviewer 1 or citations for the test set.
This truthfully reflects how we could apply our hierarchical Bayesian model in practice.

We calculate two statistics at the institutional level: the mean absolute difference (MAD) for the size-independent view and the mean absolute percentage difference (MAPD) for the size-dependent view.
These statistics are calculated for each research area (i.e. GEV) separately, using the data from the test sets.
We calculate these statistics, because we believe they are more intuitive to interpret than correlations, as also discussed by \citet{Traag2019}.
These statistics are calculated for each research area (i.e. GEV) separately.

We first focus on the size-independent view.
Let $r_i$ be the average reviewer score for institution $i$, within a specific research area.
Let there be $k$ different institutions in a specific research area.
Based on our hierarchical Bayesian model, we obtain a prediction $\hat{r}_i$ based on either one of type of citation score or the score from reviewer 1.
Note that the prediction $\hat{r}_i$ is simply defined as the average predicted reviewer score for institution $i$.
The MAD for any metric or reviewer score is then defined as
\begin{equation}
 \frac{1}{k} \sum_i \left| r_i - \hat{r}_i \right|.
\end{equation}

We now shift to the size-dependent view and the calculation of the MAPD.
Let $n_i$ denote the number of publication of institution $i$.
We then obtain the size-dependent reviewer score $n_i r_i$ and the size-dependent predicted reviewer score $n_i\hat{r}_i$.
The MAPD is then defined as
\begin{align}
 & \frac{1}{k} \sum_i \frac{\left| n_i r_i - n_i \hat{r}_i \right| }{n_i r_i} \\
 =&  \frac{1}{k} \sum_i \frac{\left| r_i - \hat{r}_i \right|}{r_i}.
\end{align}

We also calculate the MAD at the individual publication level.
The MAPD does not make sense at the individual level, as there is no size-dependent view at that level.

Note that in a Bayesian model, all estimated parameters have a posterior distribution, not just a point estimate.
That is, each estimate has some uncertainty, and this uncertainty is also reflected in any prediction $\hat{r}_i$, and hence also in the MA(P)D.

\begin{figure}[bt]
  \centering
  \includegraphics[width=\linewidth]{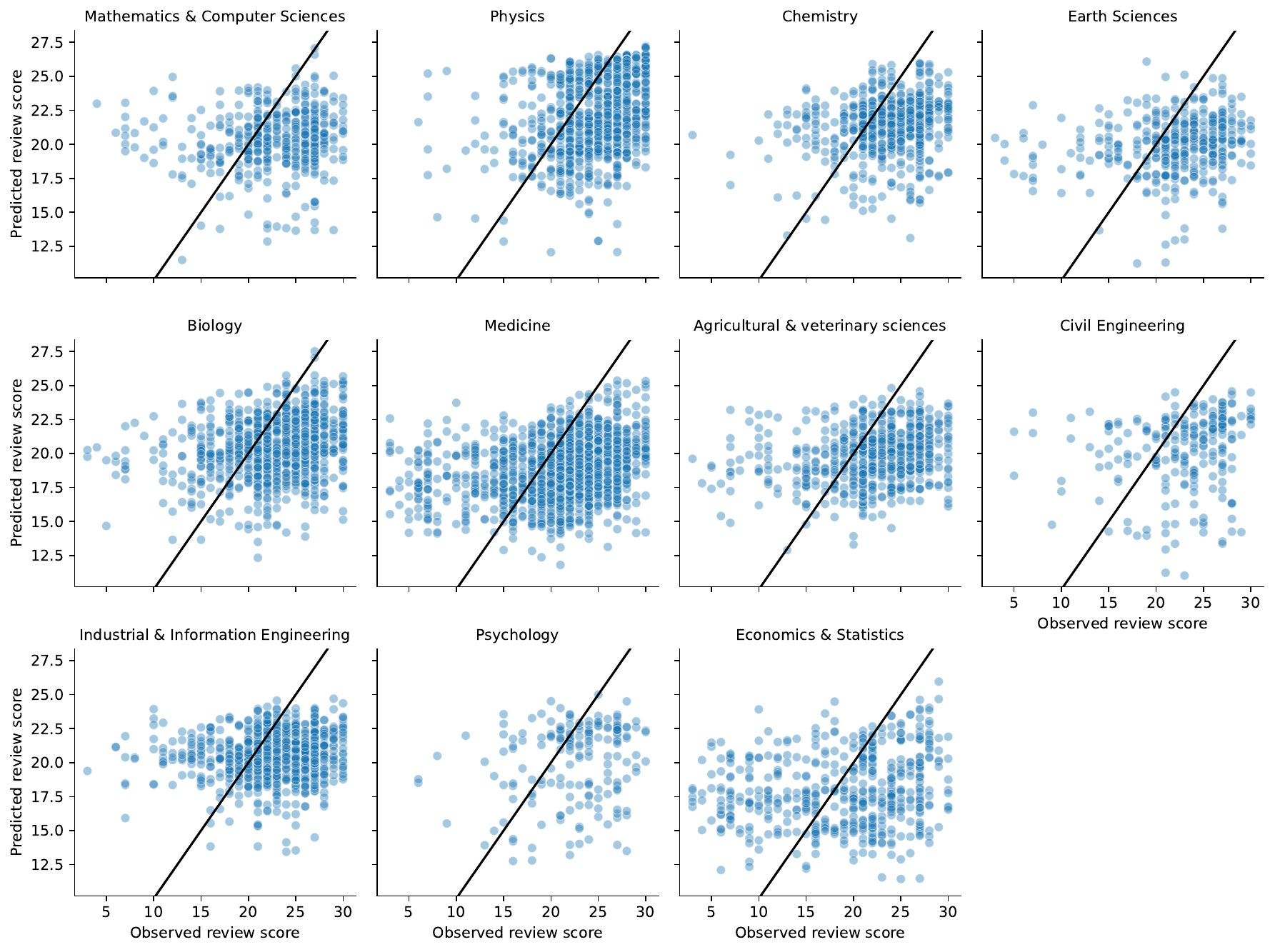}
  \caption{Scatterplots of observed reviewer scores of reviewer 2 versus mean predicted reviewer scores based on NCS.}
  \label{fig:scatter_ind_ncs}
\end{figure}

\begin{figure}[bt]
  \centering
  \includegraphics[width=\linewidth]{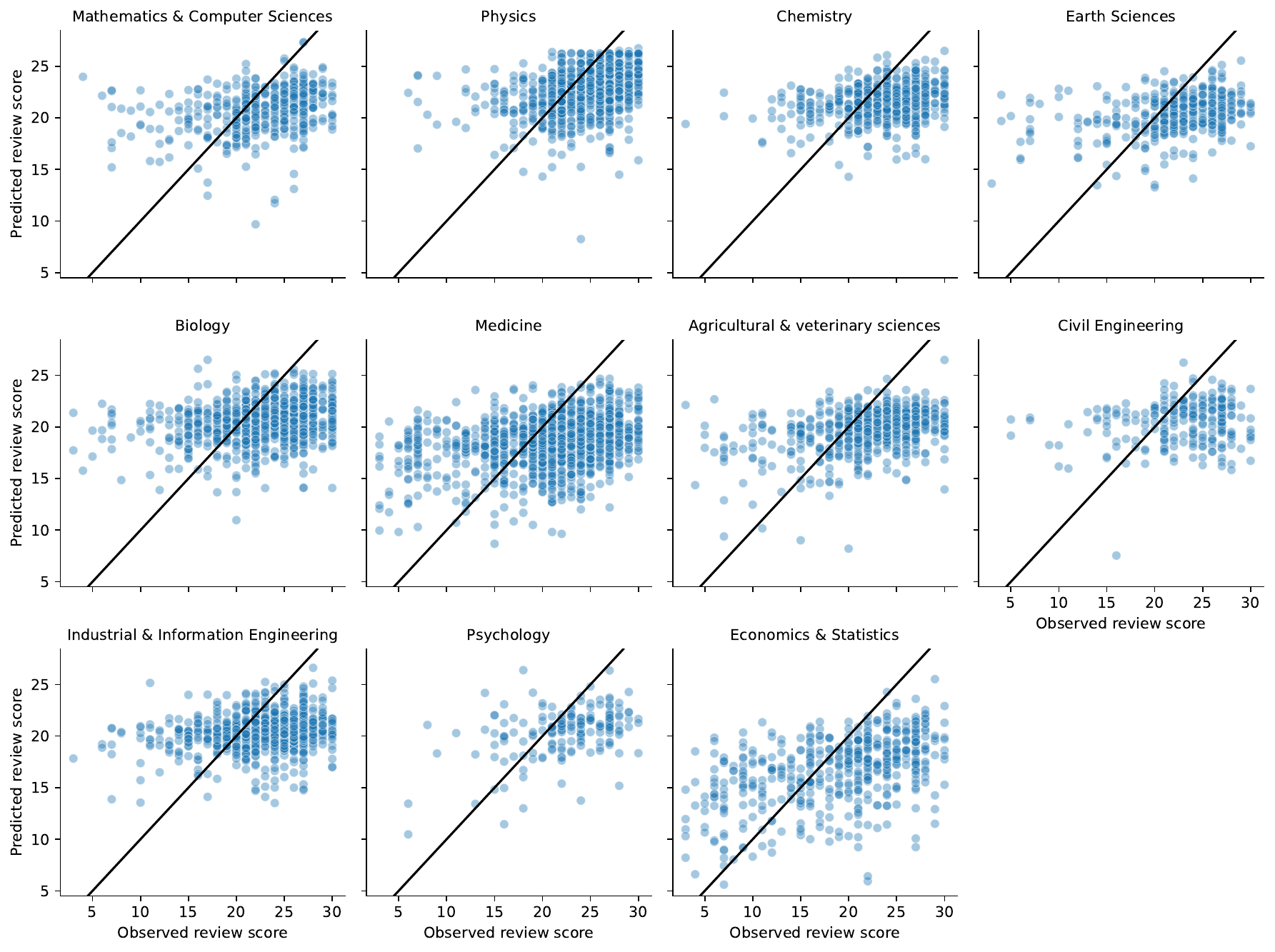}
  \caption{Scatterplots of observed reviewer scores of reviewer 2 versus mean predicted reviewer scores based on reviewer 1 at the individual level.}
  \label{fig:scatter_ind_rev}
\end{figure}

\begin{figure}[tb]
 \centering
 \includegraphics[width=\linewidth]{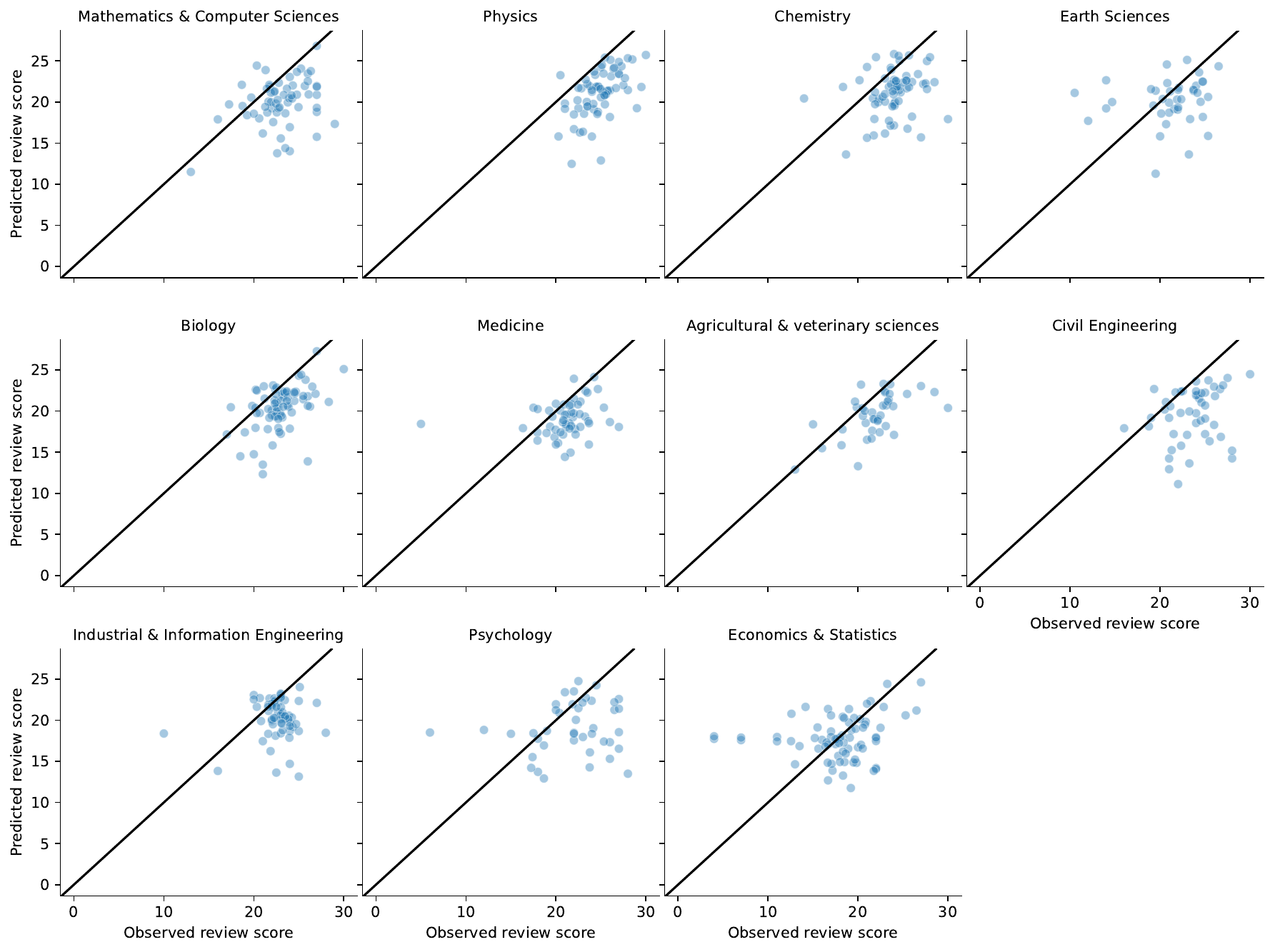}
 \caption{Scatterplots of observed reviewer scores of reviewer 2 versus predicted reviewer scores based on NCS at the institutional level, taking a size-independent view.}
 \label{fig:scatter_inst_ncs}
\end{figure}

\begin{figure}[tb]
 \centering
 \includegraphics[width=\linewidth]{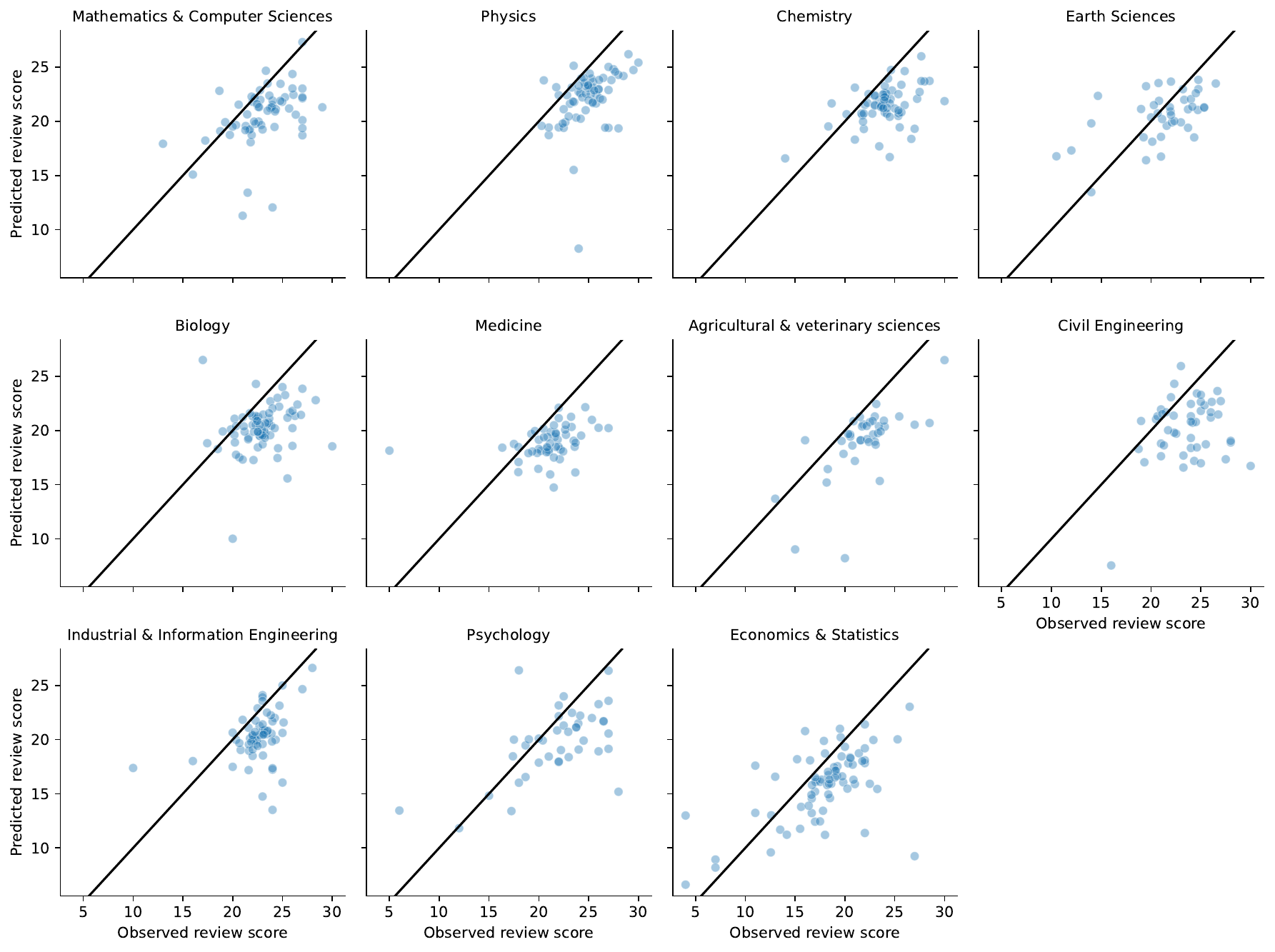}
 \caption{Scatterplots of observed reviewer scores of reviewer 2 versus predicted reviewer scores based on reviewer 1 at the institutional level, taking a size-independent view.}
 \label{fig:scatter_inst_rev}
\end{figure}

\section{Results}

\noindent In Figs.~\ref{fig:beta} and~\ref{fig:sigma} we show the posterior distribution of the various parameter estimates based on the training data.
Note that these estimates combine posterior distributions from across the $5$ different folds, resulting in some slightly visible multimodality for some parameters (e.g. the estimate for $\beta$ for NCS for GEV 2, Physics).
As is clear from Fig.~\ref{fig:beta}, the estimates from $\beta$ are more precisely estimated than the parameter estimates for predicting the non-zero probability, $\alpha_{\bar{0}}$ and $\beta_{\bar{0}}$.
Only about 10\% of NCS scores are equal to $0$, and none of the other scores.
Hence, $\alpha_{\bar{0}}$ is lower for NCS than for the other citation scores.
Since \emph{Economics \& Statistics} (GEV 13) did not list any metrics from the VQR exercise (i.e the percentile journal and percentile citation scores), these parameter estimates simply reflect the priors that we used in the model.
Overall, most $\beta$ parameters are positive, or near $0$, indicating that the citation scores are estimated to be roughly proportional to the paper value, or somewhat higher.

As illustrated in Fig.~\ref{fig:sigma}, the estimate $\sigma_\text{value}$ is quite consistent across the various citation scores.
There is some variability across various GEVs, but $\sigma_\text{value}$ is typically in the range of $0.4$--$0.5$.
Similarly, $\sigma_\text{review}$ is quite consistent across citation scores, and is typically slightly higher than $\sigma_\text{value}$, around roughly $0.6$.
There is also some consistency across different citation scores for $\sigma_\text{citation}$, but the estimates for NJS differ quite distinctly from the estimates for the other citation scores.
Interestingly, the estimates of $\sigma_\text{citation}$ are lower than the estimates of $\sigma_\text{review}$ for NJS suggesting that NJS scores show less variability for any given paper value $\phi_p$ than review scores.
Again, \emph{Economics \& Statistics} (GEV 13) essentially just shows the prior distribution for the percentile journal and percentile citation scores.

\begin{figure}
  \includegraphics[width=0.45\linewidth]{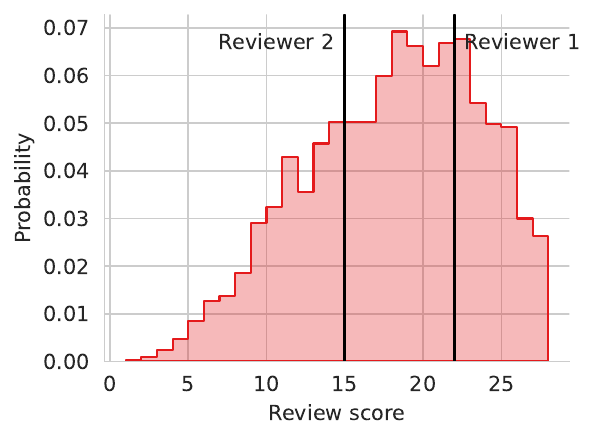}
  \includegraphics[width=0.45\linewidth]{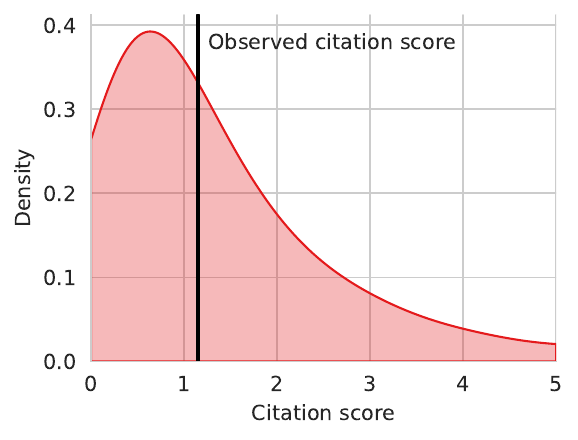}
  \caption{Posterior distributions for review scores and for the citation score (NCS) for one illustrative paper.}
  \label{fig:posterior_example}
\end{figure}

In Fig.~\ref{fig:posterior_example} we show the posterior distribution for one illustrative example paper in the training data, for both the citation score and the review scores.
This shows that the distributions can be quite broad, as already suggested in our illustration in Fig.~\ref{fig:model}.
Typically, the observed scores are within the 95\% credibility interval.
In Figs.~\ref{fig:citation_ppc} and~\ref{fig:review_ppc} we show more comprehensive posterior predictive checks for the training data.
We only show the mean predictions, and do not show the uncertainty of these predictions.
Overall, the mean predictions fit reasonably well the observations, both for citation and review scores.
There is some clear shrinkage towards the priors, especially for the percentile citation and journal scores.
The means that papers with a high score will on average be predicted to score lower, and vice versa, papers with a lower score will on average be predicted to score higher.

\begin{figure*}[]
  \centering
  \includegraphics{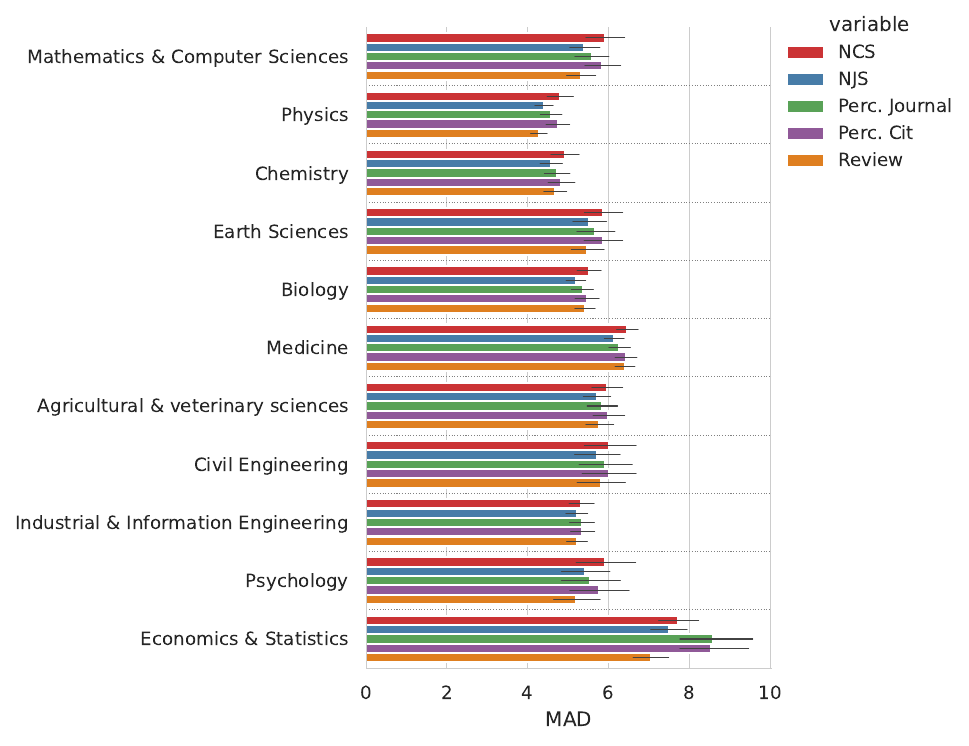}
  \caption{Mean Absolute Difference (MAD) between the score of reviewer $2$ and predictions of reviewer scores based on various bibliometric indicators and the score of reviewer $1$.
           We here calculate the MAD at the individual level.
           The error bars report the 95\% credibility interval of the posterior distribution of the MAD.}
  \label{fig:MAD_ind}
\end{figure*}

\begin{figure*}
  \centering
  \includegraphics{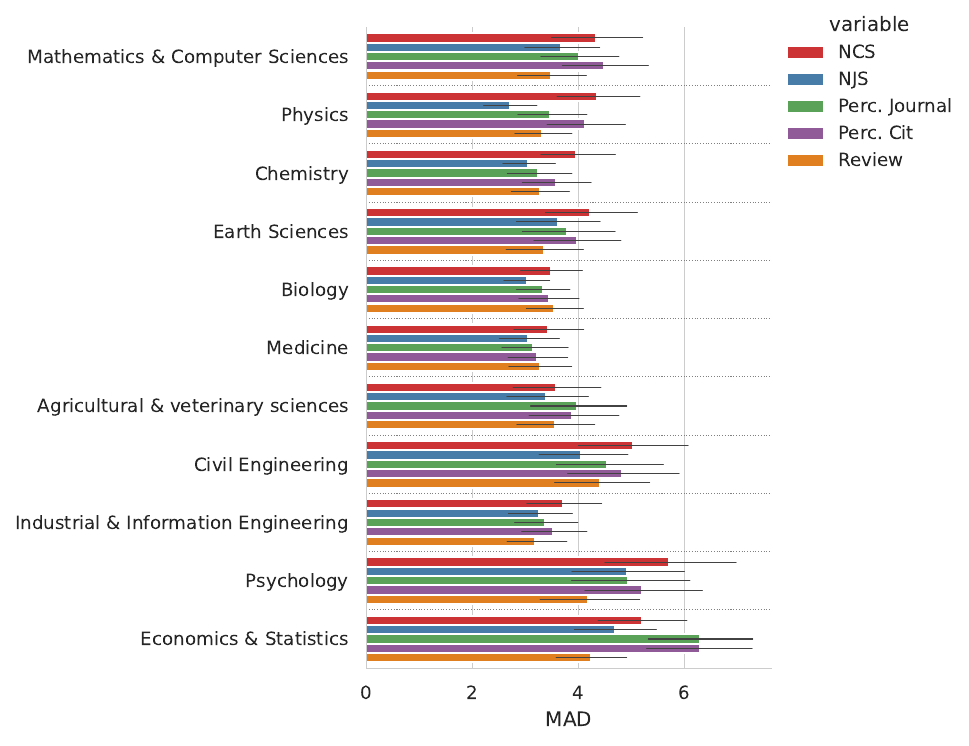}
  \caption{Mean Absolute Difference (MAD) between the score of reviewer $2$ and predictions of reviewer scores based on various bibliometric indicators and the score of reviewer $1$.
           We here calculate the MAD at the institutional level by considering the average scores for an institution, thus taking a size-independent view.
           The error bars report the 95\% credibility interval of the posterior distribution of the MAD.}
  \label{fig:MAD}
 \end{figure*}

 \begin{figure*}
  \centering
  \includegraphics{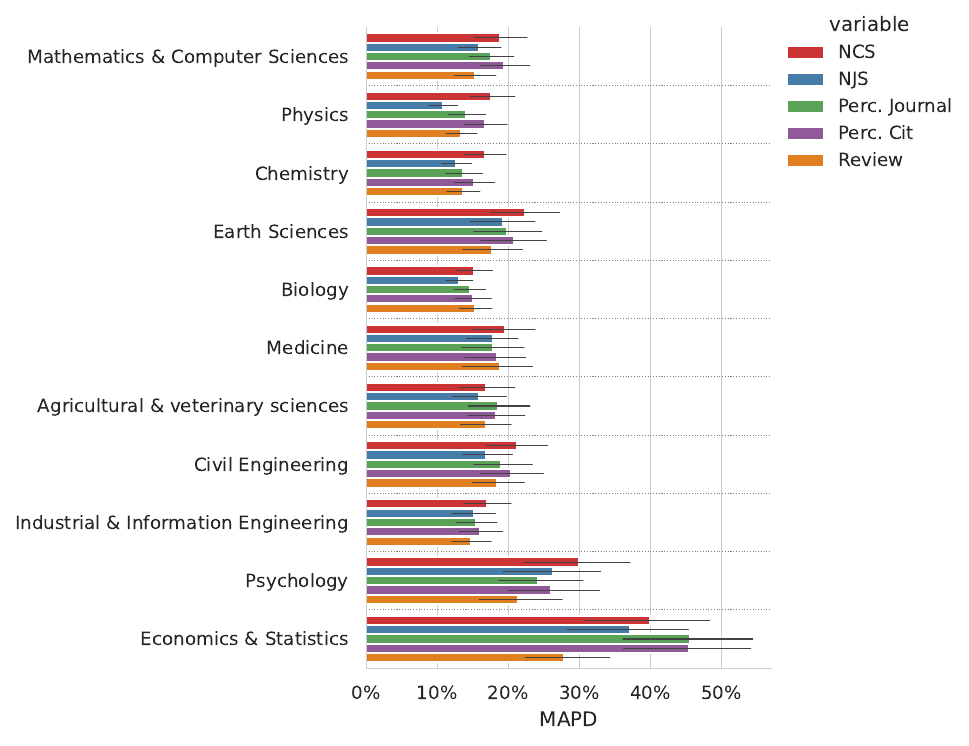}
  \caption{Mean Absolute Percentage Difference (MAPD) between the score of reviewer $2$ and predictions of reviewer scores based on various bibliometric indicators and the score of reviewer $1$.
           We here calculate the MAPD at the institutional level by considering the total scores for an institution, thus taking a size-dependent view.
           The error bars report the 95\% credibility interval of the posterior distribution of the MAPD.}
  \label{fig:MAPD}
 \end{figure*}

We now consider the results of the prediction in the test data.
We show the mean predicted reviewer score based on NCS versus the observed score of reviewer 2 in a scatter plot (Fig.~\ref{fig:scatter_ind_ncs}) and the predicted reviewer score based on the score of reviewer 1 versus the observed score of reviewer 2 in a scatter plot (Fig.~\ref{fig:scatter_ind_rev}).
It is readily apparent that there are quite some differences, not only between the NCS and peer review but also between the two reviewers themselves.
We can similarly show these results at the institutional level in Figs.~\ref{fig:scatter_inst_ncs} and~\ref{fig:scatter_inst_rev} respectively.
Although there is less variability than at the individual level, we still see subtantial differences.
By calculating the MAD and MAPD, we quantify the level of differences for each of the scores.

In Fig.~\ref{fig:avg_abs_diff_dist} we show the distribution of the absolute difference across papers.
That is, for each paper, we calculate the absolute difference between the prediction and the observed score from reviewer 2, averaged across the posterior distribution of absolute differences.
We then show the distribution of these average absolute differences over all papers.
As is clear, there are quite some individual differences between papers.
The absolute difference varies between roughly 4 and 7, but also reaches highs of 10 and abvove.
As is clear, most distributions appear quite similar both when using metrics and reviewer scores for prediction.

In Fig.~\ref{fig:MAD_ind} we report the MAD for individual publications.
Here, we calculate the absolute difference and average over all papers, that is calculate the MAD, and show the posterior distribution of the MAD.
Overall, the agreement between metrics and review is comparable to the internal agreement between two reviewers.
There are clearly some differences between research areas.
For example, the agreement is generally relatively high in Physics, while the agreement is lower in Economics \& Statistics\footnote{Note that for Economics \& Statistics, the percentile journal and percentile citation score were absent, so these scores reflect very poorly informed predictions of reviewer scores, with essentially very broad posterior prediction distribution of review scores around a mean review score of about 17.5.}.
These results also show that indicators based on citations (NCS and the citation percentile) show a similar agreement with peer review as indicators based on journal metrics (NJS and journal percentile).
Citations and journal indicators could therefore both provide information about evaluation outcomes.
The indicators based on WoS (NCS and NJS) also show a similar agreement with review scores as the more heterogeneous indicators that could be freely chosen from different data sources in the VQR (citation and journal percentiles).
This suggests that the heterogeneity of using different data sources does not deteriorate, nor ameliorate, the agreement with review.

In Fig.~\ref{fig:inst_avg_abs_diff_dist} we show the distribution of the absolute difference across institutions.
That is, it is the counterpart of Fig.~\ref{fig:avg_abs_diff_dist}, but then at the institutional level.
For each institution, we calculate the absolute difference between the prediction and the observed score from reviewer 2, averaged across the posterior distribution of absolute differences.
We then show the distribution of these average absolute differences over all institutions.
Again, most distributions appear quite similar both when using metrics and reviewer scores for prediction.
Overall, the level of differences is lower at the institutional level than at the individual level.

In Fig.~\ref{fig:MAD} and~\ref{fig:MAPD} we report the MAD and MAPD for.
Here, we calculate the absolute (percentage) difference and average over all institutions, that is calculate the MA(P)D, and show the posterior distribution of the MA(P)D.
As for individual level, the agreement between metrics and reviews at the instiutional level is again comparable to the internal agreement between two reviewers.
When comparing the MAD results at the individual publications level with the results at the institutional level (Figs.~\ref{fig:MAD_ind} vs.~\ref{fig:MAPD}), we see that the differences between research areas become less pronounced.
Overall, the MAD at the individual publication level is roughly between 4 and 6 for all indicators, including peer review itself, while at the institutional level, the MAD is roughly between 3 and 4 (Fig.~\ref{fig:MAD}).
The MAD is generally higher at the level of individual publications, compared to the institutional level, showing that ``errors'' indeed tend to ``cancel out'' at the aggregate level.
The MAPD paints a very similar picture (Fig.~\ref{fig:MAPD}) and shows an MAPD roughly between 10 and 20\%.

\section{Discussion and conclusion}

\noindent We analysed the agreement between several bibliometric indicators and peer review based on an hierarchical Bayesian model using $k$-fold cross-validation.
The contribution of our analysis is twofold: (1) we analysed the agreement at the institutional level, instead of the individual publication level; and (2) we also quantified internal reviewer agreement at the institutional level.
We found that the agreement between bibliometric indicators is on par with the internal agreement between two reviewers.
The agreement between peer review and citation-based indicators was comparable to the agreement between peer review and journal based indicators.
These results were obtained while taking into account missing review scores and missing citation scores.
Our findings are in line with findings of~\citet{Baccini2020}, who also address the issue of missing review scores.
Finally, as expected, the agreement at the institutional level was higher than at the individual publication level.

Our results are relevant in the context of performance-based research funding systems (PBRFS).
In this context, evaluations typically take place at the level of entire institutions.
Our results suggest that using peer review or bibliometric indicators, whether citation-based or journal-based, would yield similar outcomes of the evaluation.
The agreement between peer review and bibliometric indicators is not perfect, but the differences that arise are comparable to the differences between reviewers themselves.
This suggests that similarly different results could have been obtained if other reviewers would have evaluated the publications.

\begin{figure}
  \includegraphics[width=0.45\linewidth]{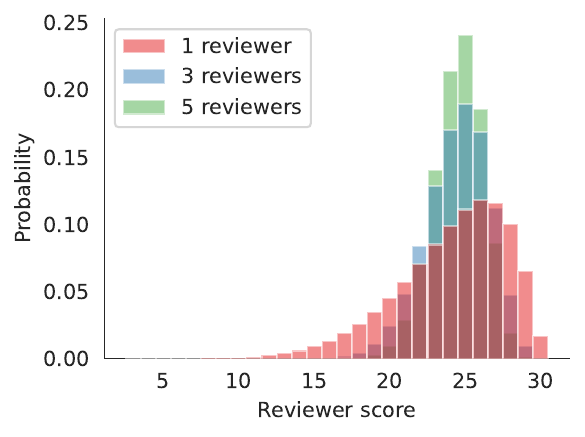}
  \includegraphics[width=0.45\linewidth]{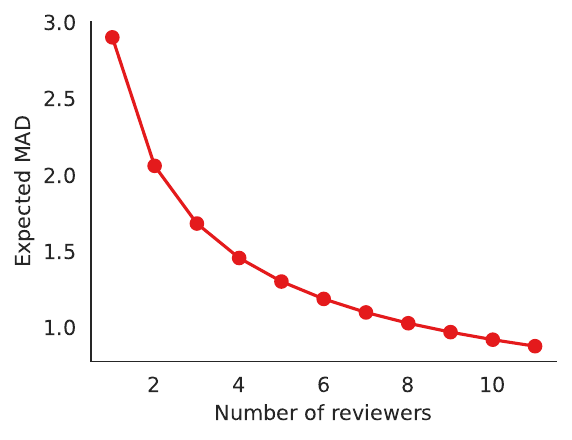}
  \caption{Illustration of the distribution of the average reviewer score when using one or more reviewers.
    The distribution of peer review scores is based on the hierarchical Bayesian model introduced earlier.
    We the same parameters as in Fig.~\ref{fig:model}, that is, $\sigma_\text{review} = 0.6$ and a paper value $\phi_p = 1.5$.
    The distribution of peer review scores using multiple reviewers is a convolution of the distribution for a single reviewer.
    The expected MAD is the expected absolute difference with the expected reviewer score.}
  \label{fig:multiple_reviewers}
\end{figure}

There are several reservations that we should make regarding our results.
First of all, our results are based on using a single reviewer for evaluation.
We may expect that using multiple reviewers will increase the internal agreement.
In reality, most evaluations do use multiple reviewers.
It is possible therefore that the agreement between bibliometric indicators and peer review is lower compared to the internal agreement when using multiple reviewers.
Based on our hierarchical Bayesian model, we can provide some insight of the internal agreement of peer review when using multiple peer reviewers.
In Fig.~\ref{fig:multiple_reviewers} we illustrate what happens with the expected MAD for a single paper.
While the expected MAD for a single paper (with these specific parameters) is almost $3$, this decreases to about $2$ for two reviewers and then to about $1.7$ and $1.5$ for additional reviewers.
Using at least two reviewers therefore seems to bring a relatively large benefit, and would be recommended, while more reviewers show diminshing marginal improvements, in terms of MAD.
To bring down the expected MAD from $3$ to $2$ requires only a single additional reviewer, but bringing it down to an MAD of $1$ requires an additional six reviewers.
This is in line with findings in the context of peer review of grant applications, where~\citet{Forscher2019-iy} suggested that as many as $12$ researchers would be needed.
Note however that this theoretical example is based on the assumption that there is a paper with a fixed value $\phi_p = 1.5$.
In reality, we would not know the paper value, and we would infer such a paper value based on the review score(s).
The uncertainty in this distribution of the paper value adds to the overall uncertainty and MAD.
For that reason, it would be of interest to conduct experiments with more than two reviewers to validate these theoretical expectations.
In addition, we now operationalise the value of a paper using a single dimension, while this could perhaps best be thought of as a multidimensional concept.
The review scores provided separate evaluations of originality, methodological rigour and impact, and teasing out the different relations with these various dimensions would be of interest.

Secondly, bibliometric indicators and peer review may show certain biases.
That is, even if the overall level of agreement is comparable, indicators and peer review may show particular differences.
For example, reviewers may be biased towards institutions of a higher reputation, and bibliometric indicators may favour certain types of methodologies or topics over others.

Thirdly, our results are obtained in the specific context of the national Italian research assessment exercise.
The assessmenmt exercise, the selection of reviewers, and the publications submitted for evaluation, are all specific to this context.
Although we expect that our results are generalisable, and are in line with earlier obervations about a low internal agreement of peer reviewers, the results may of course be different in different contexts.
Some recent evidence in the context of the UK REF quantified the internal agreement of duplicate submissions~\cite{Thelwall2022-uc}, but it is not directly clear how this compares to our results, especially because this concerns panel assessment, not independent reviewers.
It would be great if the internal agreement of post-publication peer review would also be more extensively studied in other contexts.

Additionally, there may be other effects of using indicators instead of peer review.
\citet{Rijcke2016} show that researchers may try to improve their evaluation, and they may seek to augment their bibliometric indicators, leading them to ``think with indicators'' \citep{Muller2017}, instead of pursuing scientific research as they see fit.
For example, it has been suggested that Italian authors try to improve their citation statistics by citing each other \citep{Baccini2019}.
In a seminal study, \citet{Butler2003} found that the Australian PBRFS stimulated the production of low-impact articles because only productivity was rewarded.
This result has been questioned recently by \citet{VandenBesselaar2017}, who found that impact was actually improving after the introduction of the Australian PBRFS.
Regardless of the exact findings, it remains challenging to attribute causality of the possible effects of policies in PBRFS \citep{Aagaard2017}.
Our study does not address these other possible effects.

A separate question concerns the causal mechanism that is responsible for the agreement between bibliometric indicators and peer review.
This causal mechanism is not entirely clear.
A recent study showed that citations are causally influenced by the journal \citep{Traag2021}.
Possibly, citations, journals and peer review are all influenced by common underlying characteristics of the publication.
However, it is also possible that the journal affects the peer review outcome so that a publication would have been evaluated differently if it were published elsewhere \citep{Wilsdon2015}.
Similarly, it is also possible that citations might affect the peer review outcome.
If bibliometric indicators affect the outcomes of peer review, there are a few possibilities.
On the one hand, we might want to minimize the influence of bibliometric indicators so that ``peer review'' really reflects the expert's view.
On the other hand, such an influence may be desirable and may simply reflect the fact that ``informed peer review'' has been practised.
However, if reviewers do not do much more than translating bibliometric indicators into peer review outcomes, there may also be little added benefit to peer review in this context.

In conclusion, our results show that the low agreement between a single peer reviewer and a bibliometric indicator is not necessarily worse than the internal agreement between two peer reviewers.
The low agremeement in itself is therefore no reason to reject the use of bibliometric indicators.
There are other arguments against using bibliometric indicators, such as potential biases and potential undesirable effects of using bibliometric indicators, and our results do not apply to these arguments.
In addition, when two or more independent reviewers evaluate a publication, we may expect a higher internal agreement of peer review.
Finally, our results are only applicable to a context in which a large set of publications are individually evaluated.
Other funding systems might take a different approach altogether, and forego the large-scale evaluation of publications.

\begin{acknowledgments}
  \noindent The authors thank Ludo Waltman for his earlier contributions.
\end{acknowledgments}

\section*{Conflict of interest}

\noindent The last two authors are affiliated with ANVUR, the agency tasked with executing the VQR.

\section*{Code availability}

\noindent All code necessary to replicate the results in this analysis is available from \citet{code}.

\section*{Data availability}

\noindent All data necessary to replicate the results in this analysis is available from \citet{data}.

\bibliography{bibliography}

\onecolumngrid

\appendix
\renewcommand\thefigure{S\arabic{figure}}
\setcounter{figure}{0}


\begin{figure*}
  \includegraphics[width=\linewidth]{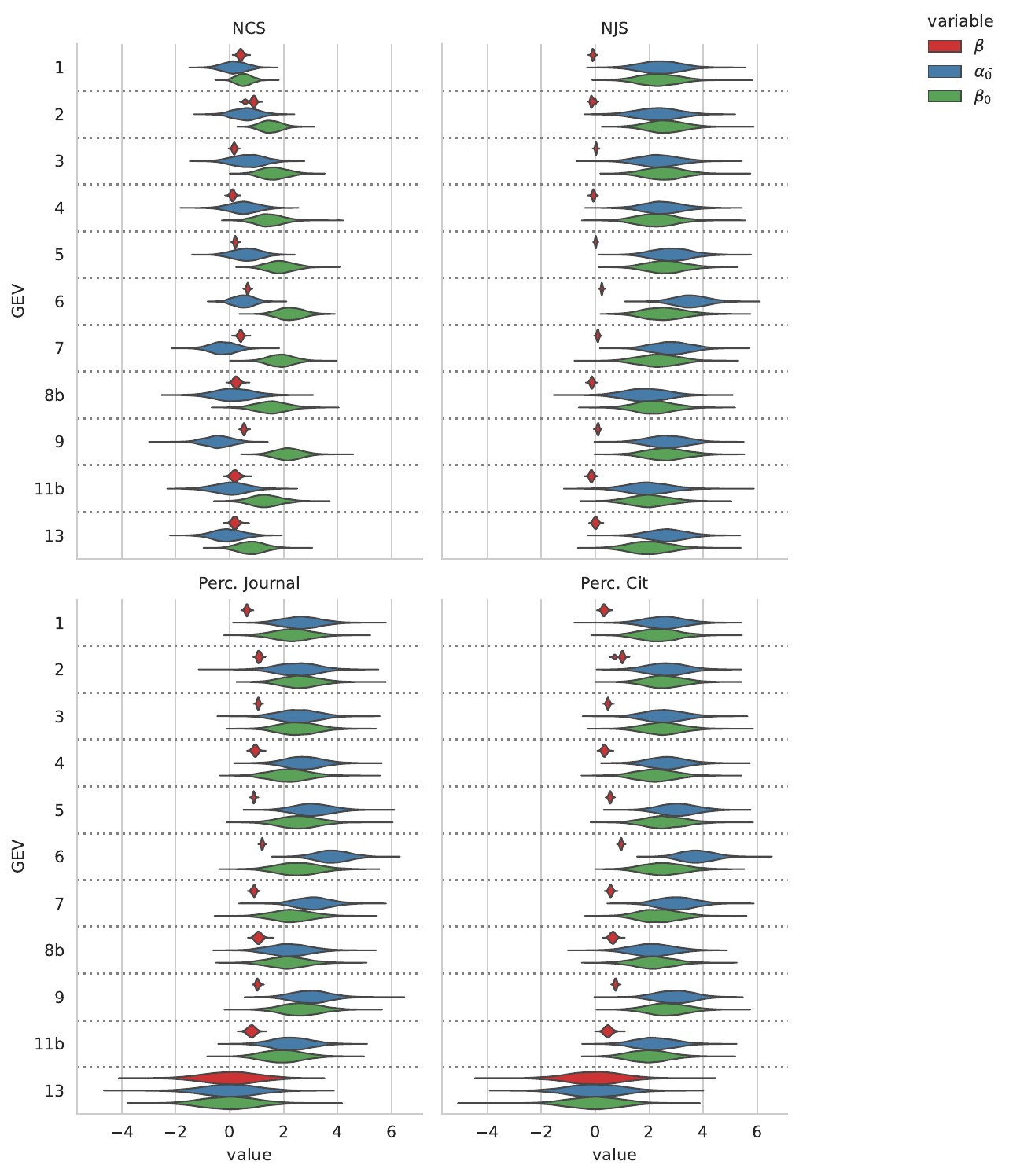}
  \caption{Parameter estimates of $\beta$, $\alpha_{\bar{0}}$ and $\beta_{\bar{0}}$ for the various GEV and citation scores.}
  \label{fig:beta}
\end{figure*}

\begin{figure*}
  \includegraphics[width=\linewidth]{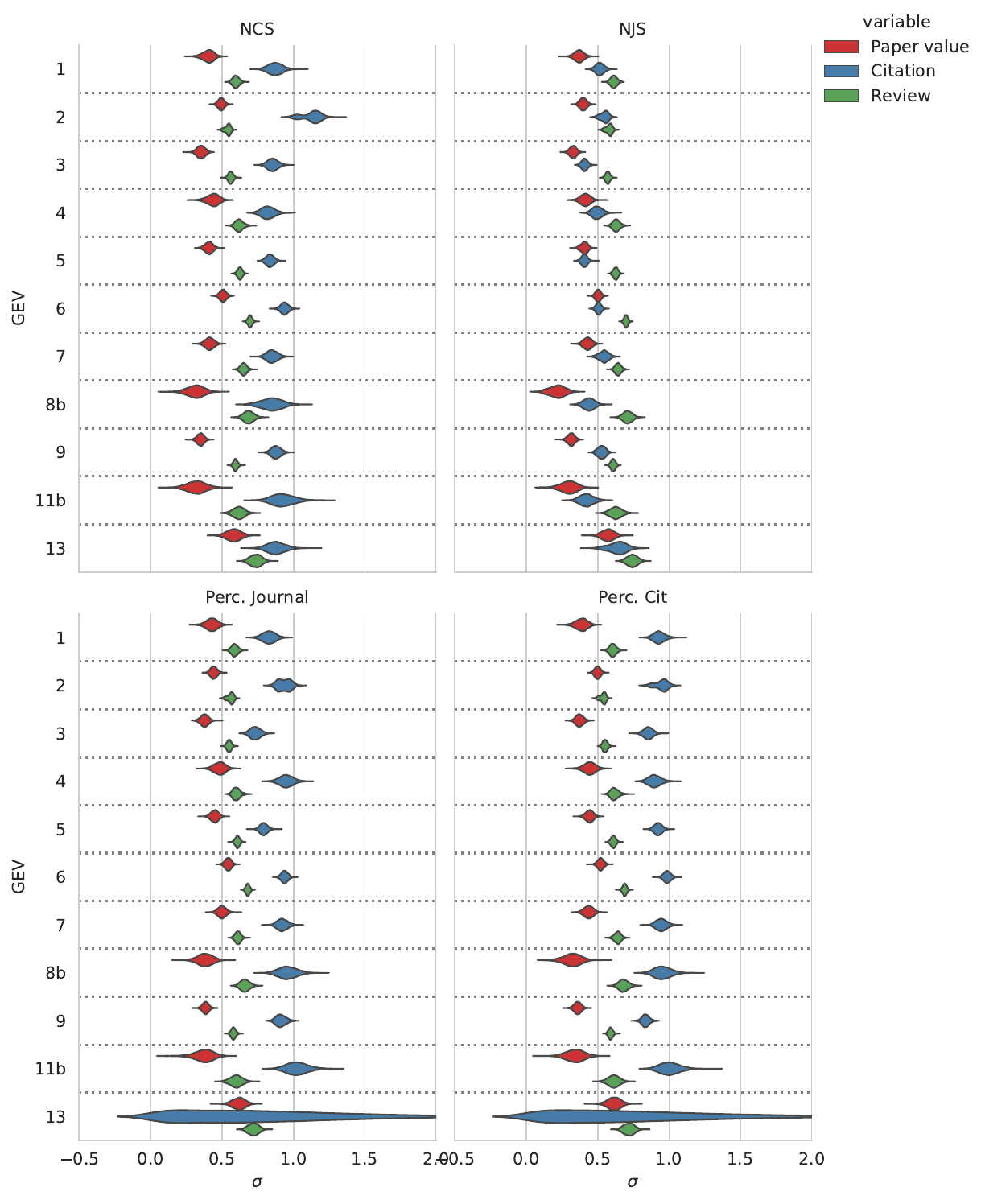}
  \caption{Parameter estimates $\sigma_\text{value}$, $\sigma_\text{citation}$ and $\sigma_\text{review}$ for the various GEV and citation scores.}
  \label{fig:sigma}
\end{figure*}


\begin{figure*}
  \centering
  \includegraphics{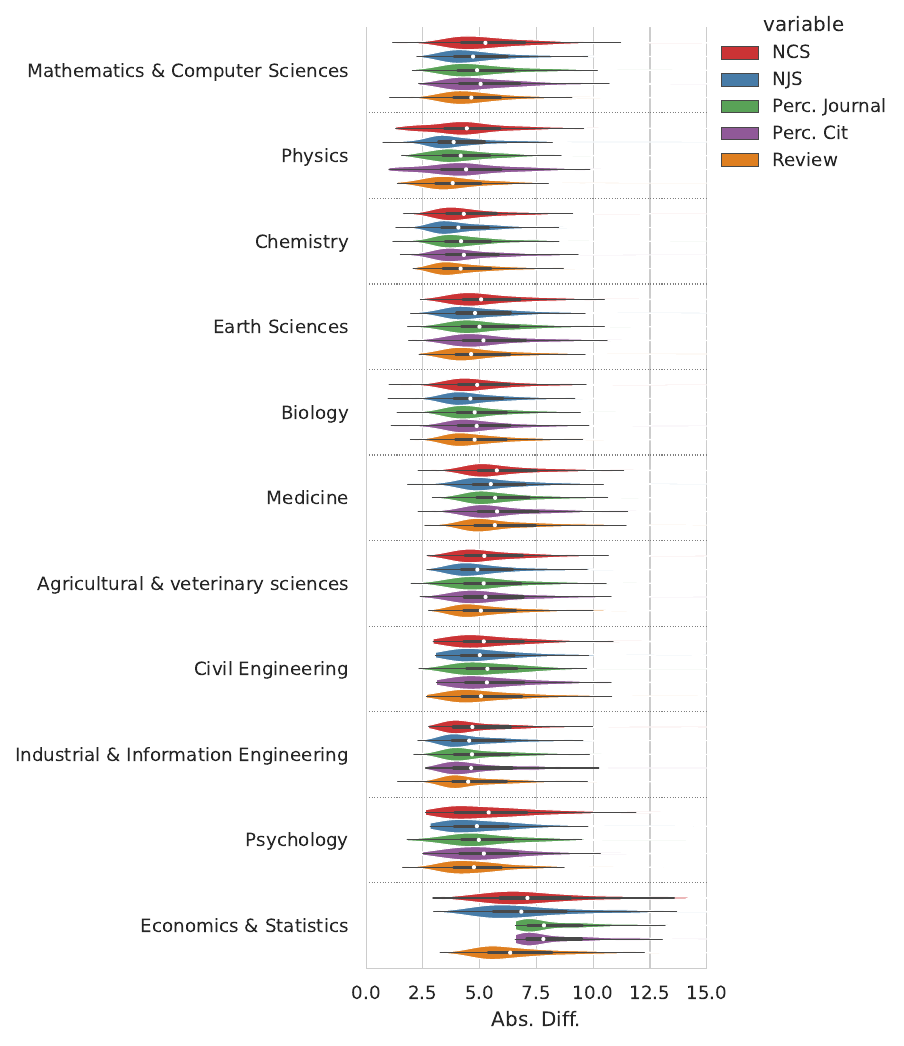}
  \caption{Distribution of the absolute differences across papers.
           We here calculate the absolute difference for each paper, averaged across the posterior distribution of absolute differences, and show the distribution of the average absolute differences.
           }
  \label{fig:avg_abs_diff_dist}
\end{figure*}

\begin{figure*}
  \centering
  \includegraphics{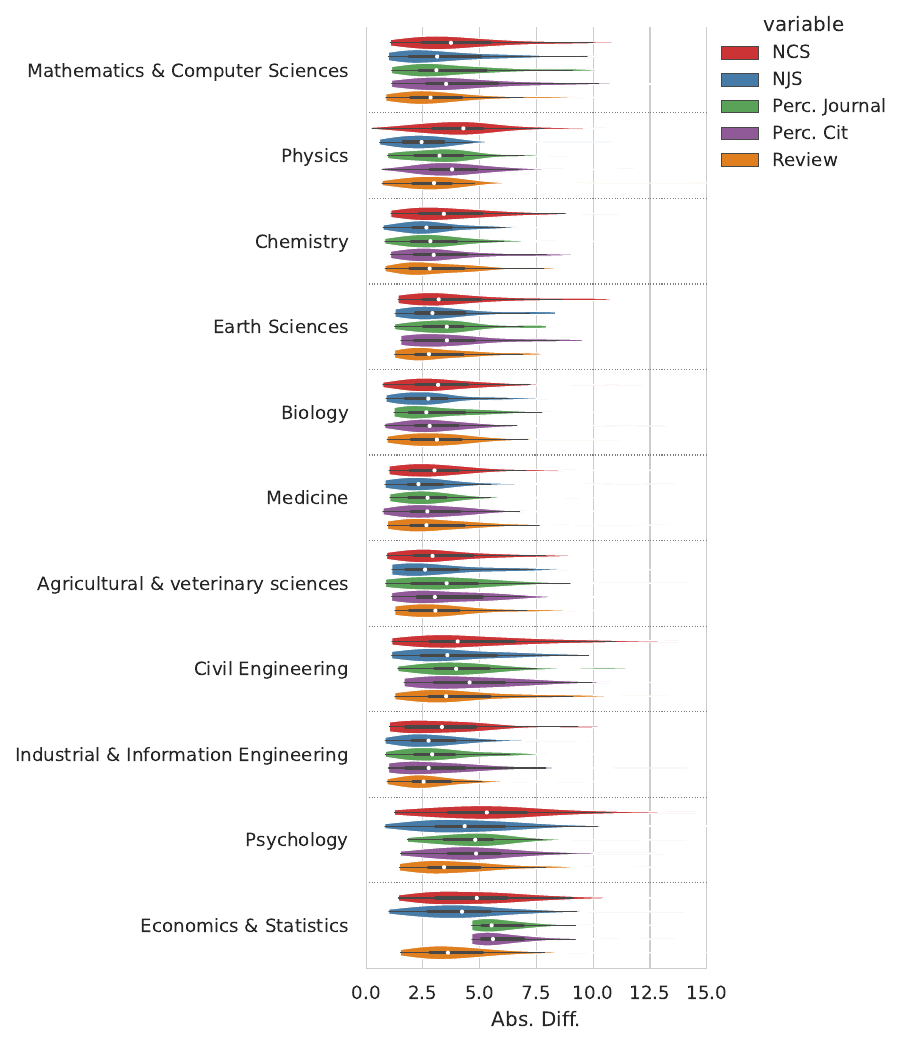}
  \caption{Distribution of the absolute differences across institutes.
           We here calculate the absolute difference for each institute, averaged across all draws from the posterior distribution, and show the distribution of the average absolute differences.
           }
  \label{fig:inst_avg_abs_diff_dist}
\end{figure*}


\begin{figure}
  \includegraphics[width=0.45\linewidth]{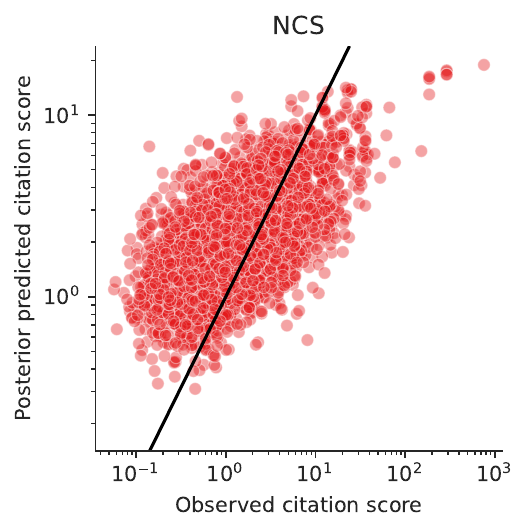}
  \includegraphics[width=0.45\linewidth]{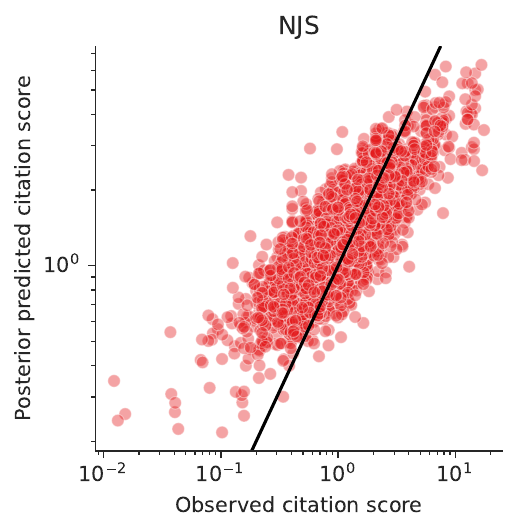}
  \includegraphics[width=0.45\linewidth]{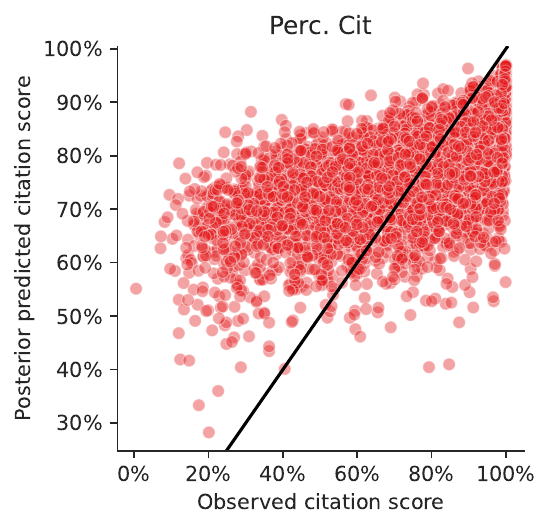}
  \includegraphics[width=0.45\linewidth]{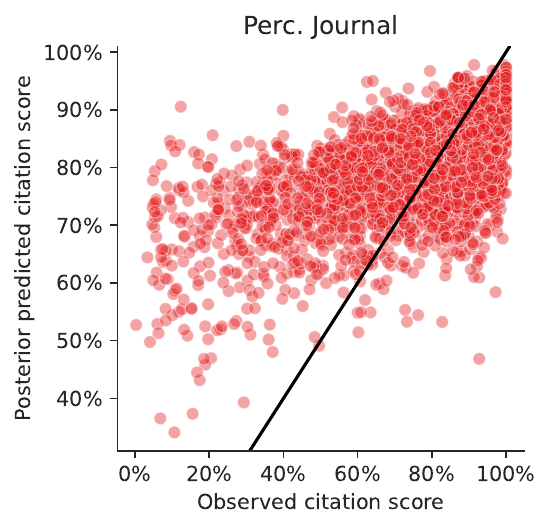}
  \caption{Posterior predictive check of citation scores.}
  \label{fig:citation_ppc}
\end{figure}

\begin{figure}
  \includegraphics[width=0.45\linewidth]{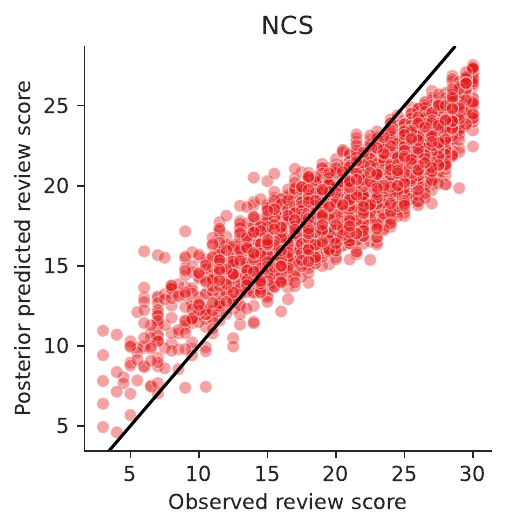}
  \includegraphics[width=0.45\linewidth]{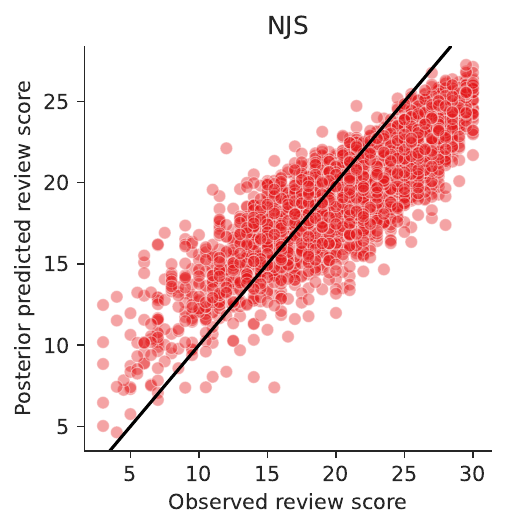}
  \includegraphics[width=0.45\linewidth]{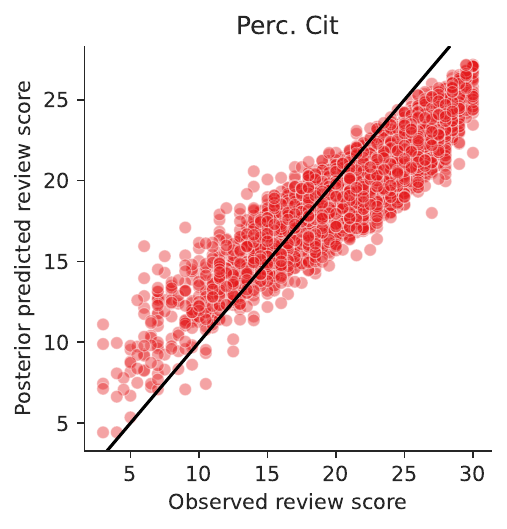}
  \includegraphics[width=0.45\linewidth]{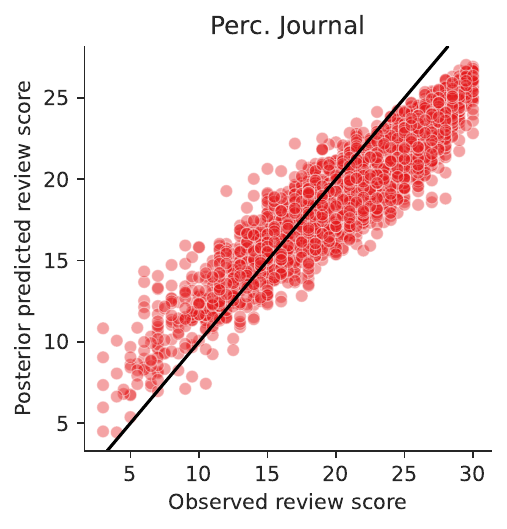}
  \caption{Posterior predictive check of review scores.}
  \label{fig:review_ppc}
\end{figure}

\end{document}